\documentclass{webofc}
\usepackage[varg]{txfonts} % Web of Conferences font
\usepackage{lineno}
\begin{document}
\title{Production of pions, kaons, and (anti-)protons in Au+Au collisions at $\sqrt{s_{NN}}$ = 54.4 GeV at RHIC}

% subtitle is optional

\author{\firstname{Krishan}
\lastname{Gopal$^{1}$} 
\fnsep\thanks{\email{krishangopal@students.iisertirupati.ac.in}
}, for the STAR collaboration
}

\institute{Indian Institute of Science Education and Research (IISER) Tirupati, India}

\abstract{%
We report systematic measurements of bulk properties of the system created in Au+Au collisions at $\sqrt{s_{NN}}$ = 54.4 GeV recorded by the STAR detector at the Relativistic Heavy Ion Collider (RHIC). The transverse momentum spectra of $\pi^{\pm}$, $ K^{\pm}$, and $p$($\bar{p}$) are studied at mid-rapidity $(|y| < 0.1)$ for different centrality classes. We also present mid-rapidity measurement of particle yields $(dN/dy)$ and particle ratios in Au+Au collisions at $\sqrt{s_{NN}}$ = 54.4 GeV. The kinetic freeze-out parameters ($T_{kin}$ and $⟨\beta⟩$) are extracted from particle spectra using the blast-wave model, and all the results are compared with previously published experimental results.% as a function of collision centrality and energy.
  
  
  }
\maketitle
\section{Introduction}
\label{intro}
Quantum Chromodynamics (QCD) predicts that at very high energy density and/or very high temperature, a new state of matter is formed where quarks and gluons are deconfined, known as the Quark-Gluon Plasma (QGP) \cite{RefC}. The study of transverse momentum spectra of produced particles is important for investigating the bulk properties of the QGP, such as integrated yields $(dN/dy)$, average transverse momentum $\langle p_{T} \rangle$, particle ratios, and freeze-out properties. This information is also useful for understanding particle production mechanisms. 
\section{Analysis details}
We have studied the transverse momentum spectra of $\pi^{\pm}$, $K^{\pm}$, and $p$($\bar{p}$) at mid-rapidity $(|y| < 0.1)$ for different centrality classes in Au+Au collisions at $\sqrt{s_{NN}}$ = 54.4 GeV, recorded by STAR \cite{star} experiment in the year 2017. In the STAR experiment, charged particle identification is performed utilizing the Time Projection Chamber (TPC) and Time of Flight (TOF) detectors depending on the momentum of the particle.  
For low momentum particles, the TPC is used, while for particles with intermediate or high momenta ($p_{T}$ > 1 GeV/c), the TOF is used. TPC works on the principle of ionization energy loss of the charged particles passing through TPC's gas volume. \\
The TPC uses the ionization energy loss ($dE/dx$) for particle identification, from which the variable $z$ is defined:
\begin{align*}
  z_{X} =  ln(\frac{<dE/dx>}{<dE/dx>^{B}_{X}})\\
 \end{align*}
%\newpage
where $<dE/dx>^{B}_{X}$ is the expected energy loss based on the Bichsel function [3] and $X$ is the particle type ($e^{\pm}$, $\pi^{\pm}$, $K^{\pm}$, $p$,  or  $\bar{p}$).
A multi-Gaussian fit is carried out on $z_{X}$ distributions to extract raw yields. \\
For the TOF detector, we use mass square information, $m^{2} = p^{2} (\frac{c^{2}T^{2}}{L^{2}} -1)$ where $m$,  $p$,  $L$,  $T$ and $c$ are mass of particle,  momentum of the particle,   path length, time of travel by the particle and speed of light, respectively. From the above two methods we obtain raw spectra from TPC and TOF. The TPC tracking efficiency and acceptance,  energy loss correction,  pion and proton background subtraction are applied to calculate corrected spectra. For the TOF analysis, TOF matching efficiency correction is also applied. 
\vspace{-0.4cm}
\section{Results and discussions}
\subsection{Transverse momentum spectra and particle yields}
We calculate the $p_{T} $ spectra of $\pi^{\pm}$, $K^{\pm}$, and  $p$($\bar{p}$) in Au+Au collisions at $\sqrt{s_{NN}}$ = 54.4 GeV. 
Figure 1 shows the $p_{T}$ spectra of $\pi^{+}$, $K^{+}$, and $p$ in different centrality classes from central (0-5\%) to peripheral (70-80\%) collisions. The $p_{T}$ spectra of identified hadrons show a clear centrality dependence. We fit the pion ($\pi^{\pm}$), kaon ($K^{\pm}$) spectra with the Levy function and $p$($\bar{p}$) spectra with a double-exponential function, to calculate particle yields ($dN/dy$). 
%\begin{center}
\begin{figure}[h!]
\centering
\includegraphics[width=11cm]{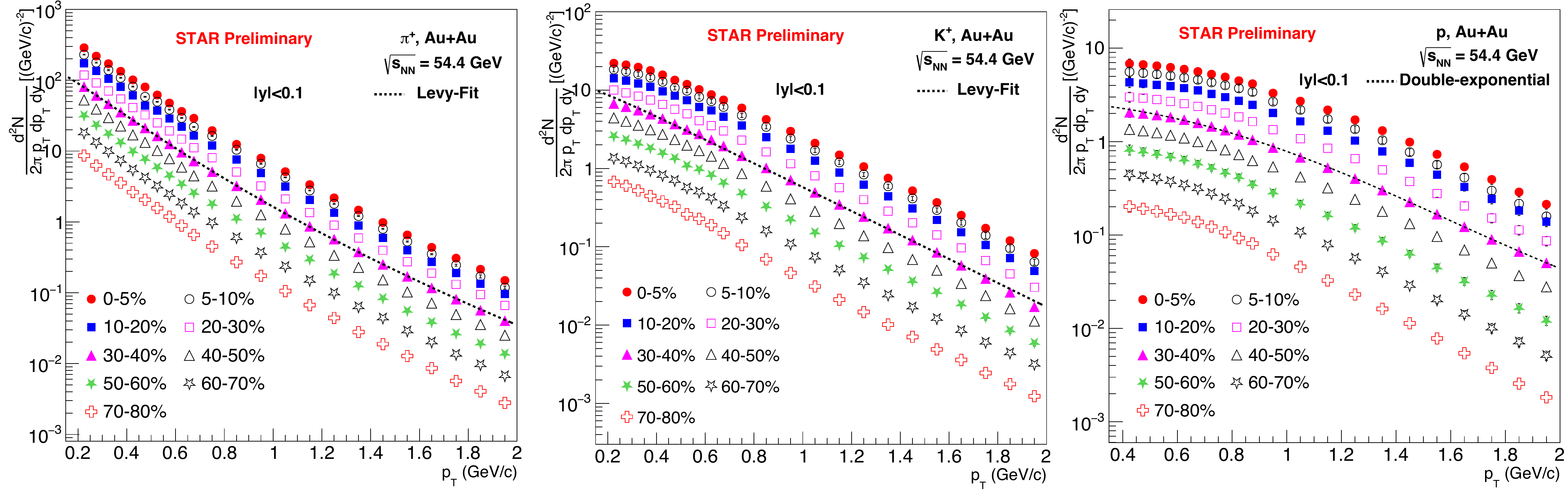}
\vspace{-0.2cm}
\caption{$p_{T}$ spectra of $\pi^{+}$, $K^{+}$, and $p$ measured at mid-rapidity ($|y|$ $<$ 0.1) in Au+Au collisions at $\sqrt{s_{NN}}$ = 54.4 GeV in different collision centralities. The statistical and systematic uncertainties are added in quadrature.}
\label{fig:Fig.1}
\end{figure}
%\newpage
\vspace{-0.7cm}
\begin{figure}[h!]
\centering
\includegraphics[width=11cm]{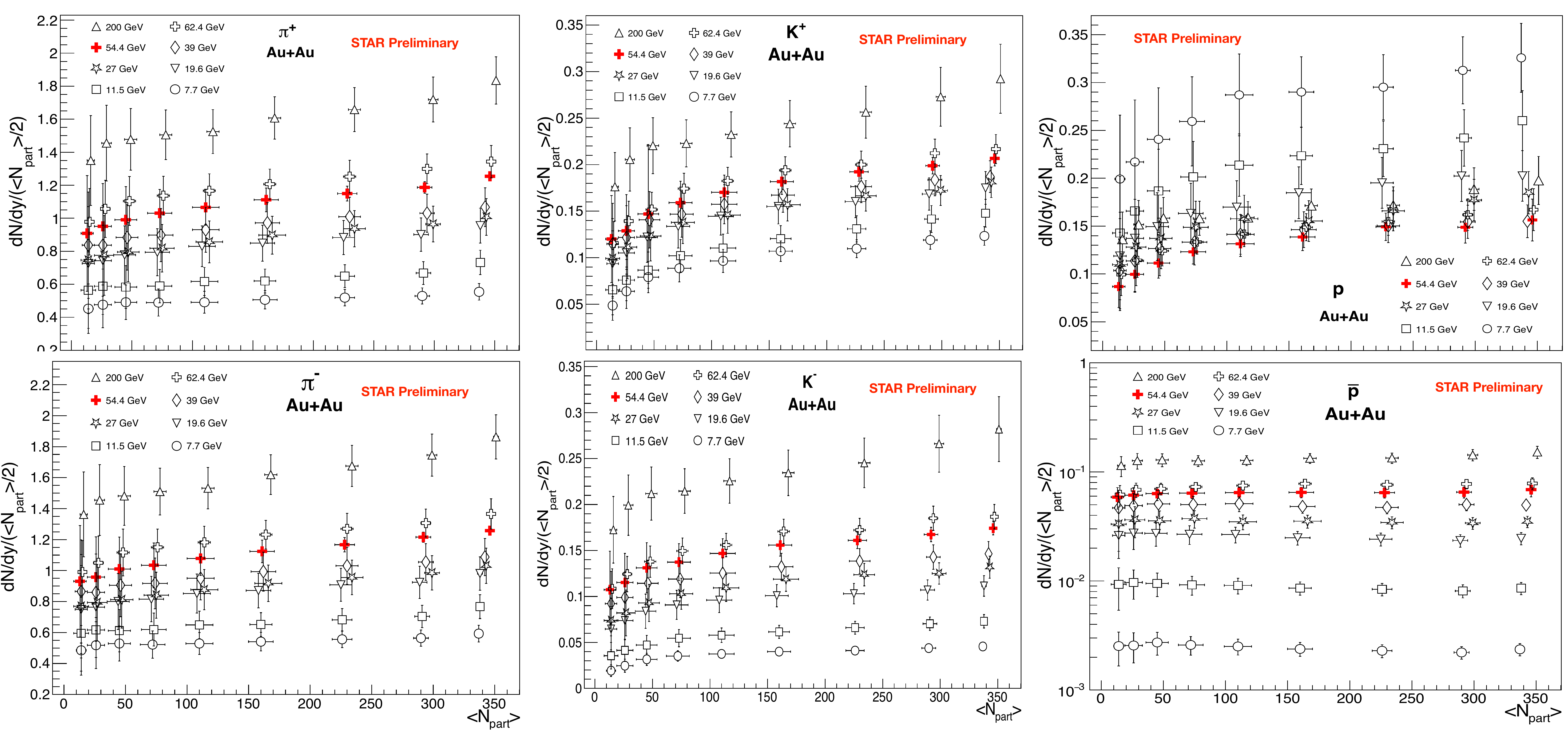}
\vspace{-0.3cm}
\caption{Centrality dependence of $dN/dy$ normalized by $\langle N_{\text{part}} \rangle/2$ for $\pi^{\pm}$, $ K^{\pm}$, and $p$($\bar{p}$) at mid-rapidity ($|y| < 0.1$) in Au+Au collisions at $\sqrt{s_{NN}}$ = 54.4 GeV. Errors shown are quadrature sums of statistical and systematic uncertainties.}
\label{fig:Fig.2}
\end{figure}
\vspace{-0.6cm}
Figure 2 shows the centrality dependence of $dN/dy$  normalized by half of the average number of participating nucleons ($\langle N_{\text{part}} \rangle$/2) for $\pi^{\pm}$, $ K^{\pm}$, and $p$($\bar{p}$) in Au+Au collisions at $\sqrt{s_{NN}}$ = 54.4 GeV, along with other RHIC's Beam Energy Scan (BES) energies. \\
The yields of $\pi^{\pm}$, $K^{\pm}$, and $\bar{p}$ decrease with decreasing collision energy, whereas the proton yield is the highest for the lowest energy of 7.7 GeV, which indicates the largest baryon stopping at mid-rapidity at this energy.
\vspace{-0.4cm}
\subsection{Particle ratios}
\vspace{-0.4cm}
\begin{figure}[h!]
\centering
\includegraphics[width=12cm]{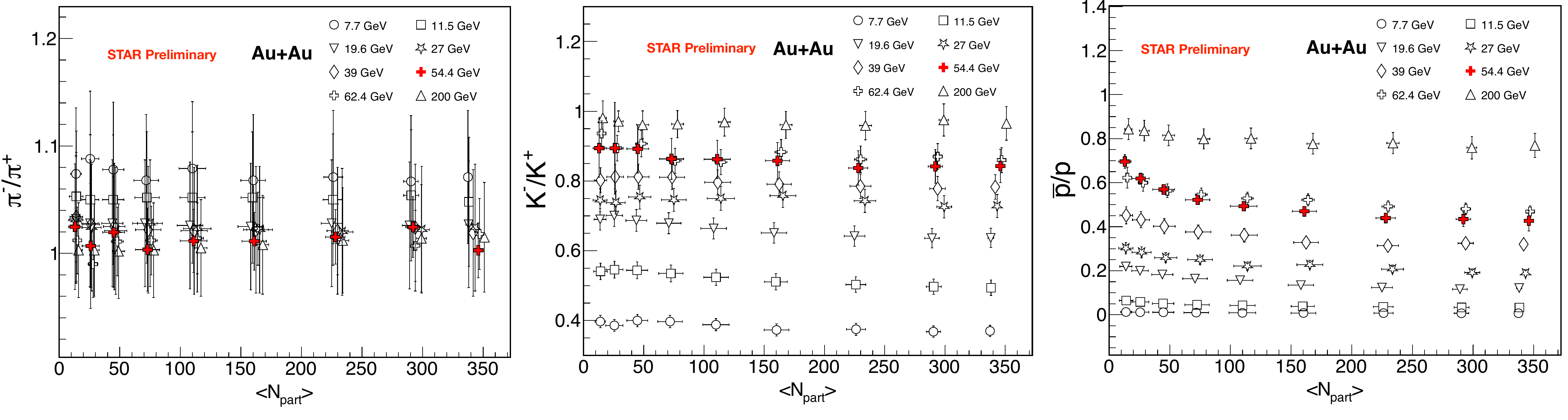}
\vspace{-0.3cm}
    \caption{Variation of $\pi^{-}/\pi^{+}$, $K^{-}/K^{+}$, and $\bar{p}/p$ ratios as a function of $\langle N_{\text{part}} \rangle$ at mid-rapidity ($|y| < 0.1$) in Au+Au collisions at $\sqrt{s_{NN}}$ = 54.4 GeV.}
    \label{fig:Fig.3}
\end{figure}
\vspace{-0.65cm}
 Particle ratios of $\pi^{\pm}$, $ K^{\pm}$, and $p$($\bar{p}$) in Au+Au collisions at $\sqrt{s_{NN}}$ = 54.4 GeV are shown in Fig. 3. The $\pi^{-}/\pi^{+}$ ratio is close to unity. The $K^{-}/K^{+}$ ratio shows a mild centrality dependence. %The $\bar{p}/p$ ratio increases from central to peripheral collisions. 
 The $\bar{p}/p$ ratio decreases from peripheral to central collisions. Figure 4 shows the centrality dependence of mixed ratios ($K^{-}/\pi^{-}$, $K^{+}/\pi^{+}$, $\bar{p}/\pi^{-}$, and $p/\pi^{+}$). With increasing energy $K^{-}/\pi^{-}$ shows an increasing behavior. The $K^{+}/\pi^{+}$ ratio has the maximum value at 7.7 GeV due to the associate production of $K^{+}$ at low energy. Also $K^{+}/\pi^{+}$ increases from peripheral to central collisions. The $\bar{p}/\pi^{-}$ ratio shows an increasing trend with increasing collision energy and shows a mild centrality dependence. The $p/\pi^{+}$ ratio decreases with increasing energy due to reduced baryon stopping as we go to higher center-of-mass energies.
 \vspace{-0.3cm}
\begin{figure}[h!]
    \centering
   \includegraphics[width=8cm]{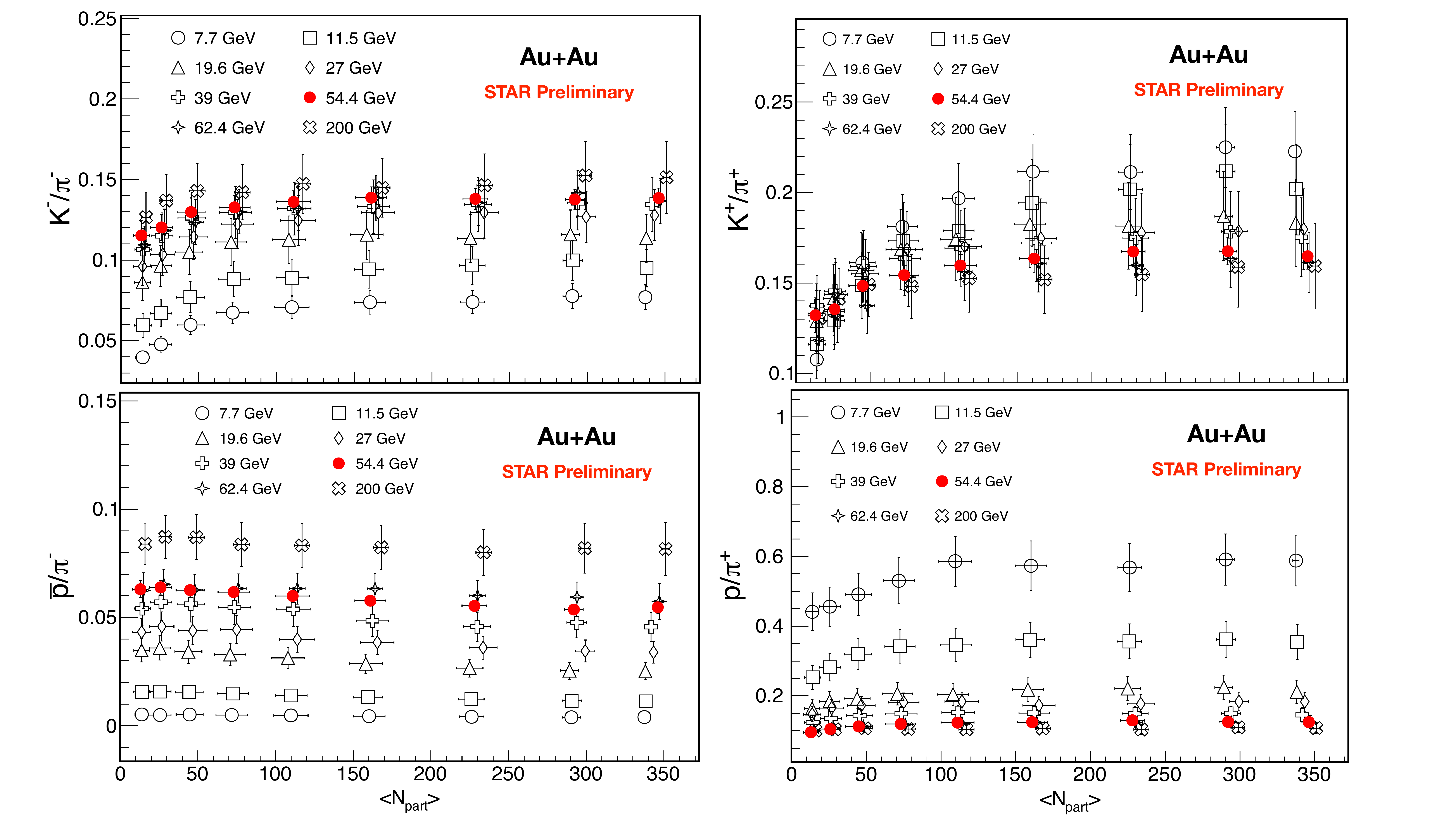}
   \vspace{-0.3cm}
    \caption{$K^{-}/\pi^{-}$, $K^{+}/\pi^{+}$, $\bar{p}/\pi^{-}$,  and $p/\pi^{+}$ ratios as a function of $\langle N_{\text{part}} \rangle$ at mid-rapidity ($|y| < 0.1$) in Au+Au collisions at $\sqrt{s_{NN}}$ = 54.4 GeV.}
    \label{fig:Fig.4}
\end{figure}
\vspace{-0.7cm}
\begin{figure}[h!]
    \centering
    \includegraphics[width=8cm]{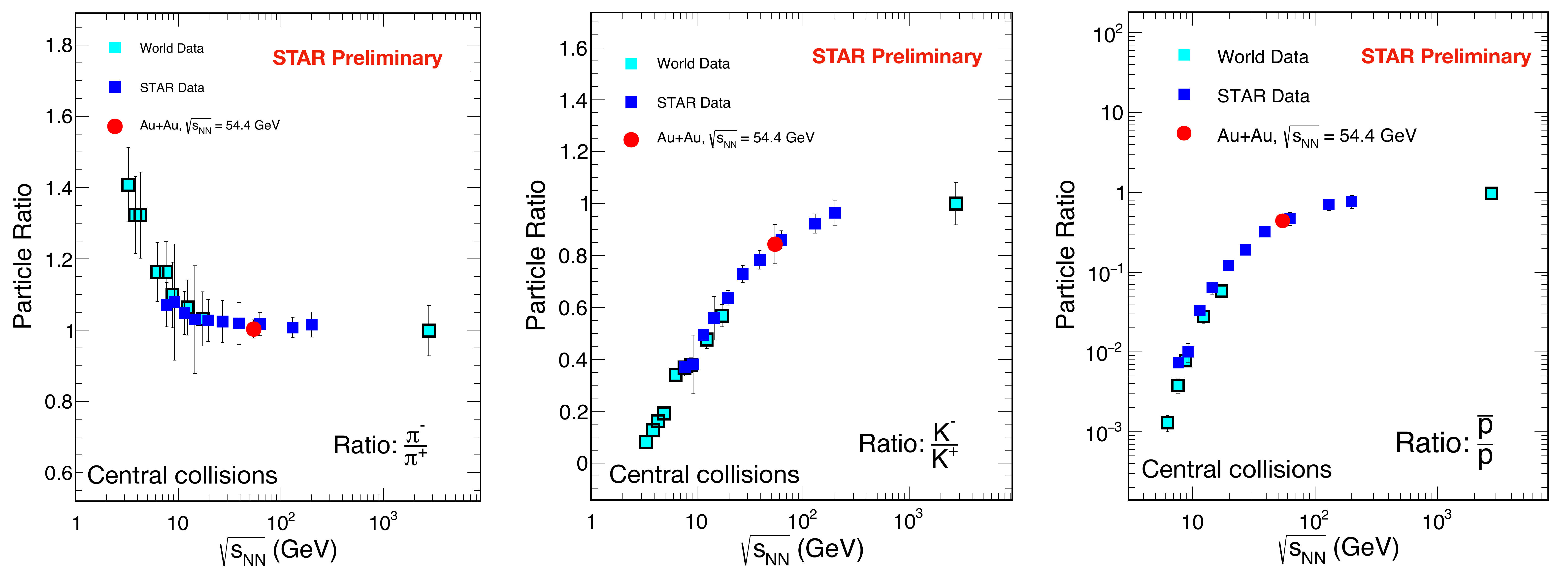}
    \vspace{-0.4cm}
    \caption{$\pi^{-}/\pi^{+}$, $K^{-}/K^{+}$, and $\bar{p}/p$ ratios at mid-rapidity ($|y| < 0.1$) in 0–5\% Au+Au collisions at $\sqrt{s_{NN}}$ from 7.7-200 GeV.}
    \label{fig:Fig.5}
\end{figure}
\vspace{-0.5cm}
%\newpage
Figure 5 shows the collision energy dependence of the particle ratios $\pi^{-}/\pi^{+}$, $K^{-}/K^{+}$, and $\bar{p}/p$ in most central collisions at different energies.
The latest result from 54.4 GeV follows the trend shown from previous measurements of AGS, SPS, RHIC, and LHC \cite{RefJ} \cite{RefD}. The displayed uncertainties are statistical and systematic uncertainties added in quadrature for each ratio plots (Fig. 3, Fig. 4, and Fig. 5).
\subsection{Kinetic freeze-out parameters}
The left panel of Fig. 6 shows the simultaneous fit to the $p_T$ spectra of $\pi$, K, p and its anti-particle in 0-5\% central collisions in $\sqrt{s_{NN}}$ = 54.4 GeV collisions using the blast-wave model \cite{blastwave} \cite{blastwave2}. The right panel of Fig. 6 shows the variation of kinetic freeze-out temperature ($T_{kin}$) with average transverse radial flow velocity ($\langle \beta \rangle$) for the different center-of-mass energies and centralities.  $\langle \beta \rangle$ decreases as we go from most central to peripheral collisions, whereas $T_{kin}$ increases from central to peripheral collisions due to the short-lived fireball formed in peripheral collisions.
\begin{figure}[h!]
    \centering
   \includegraphics[width=9cm]{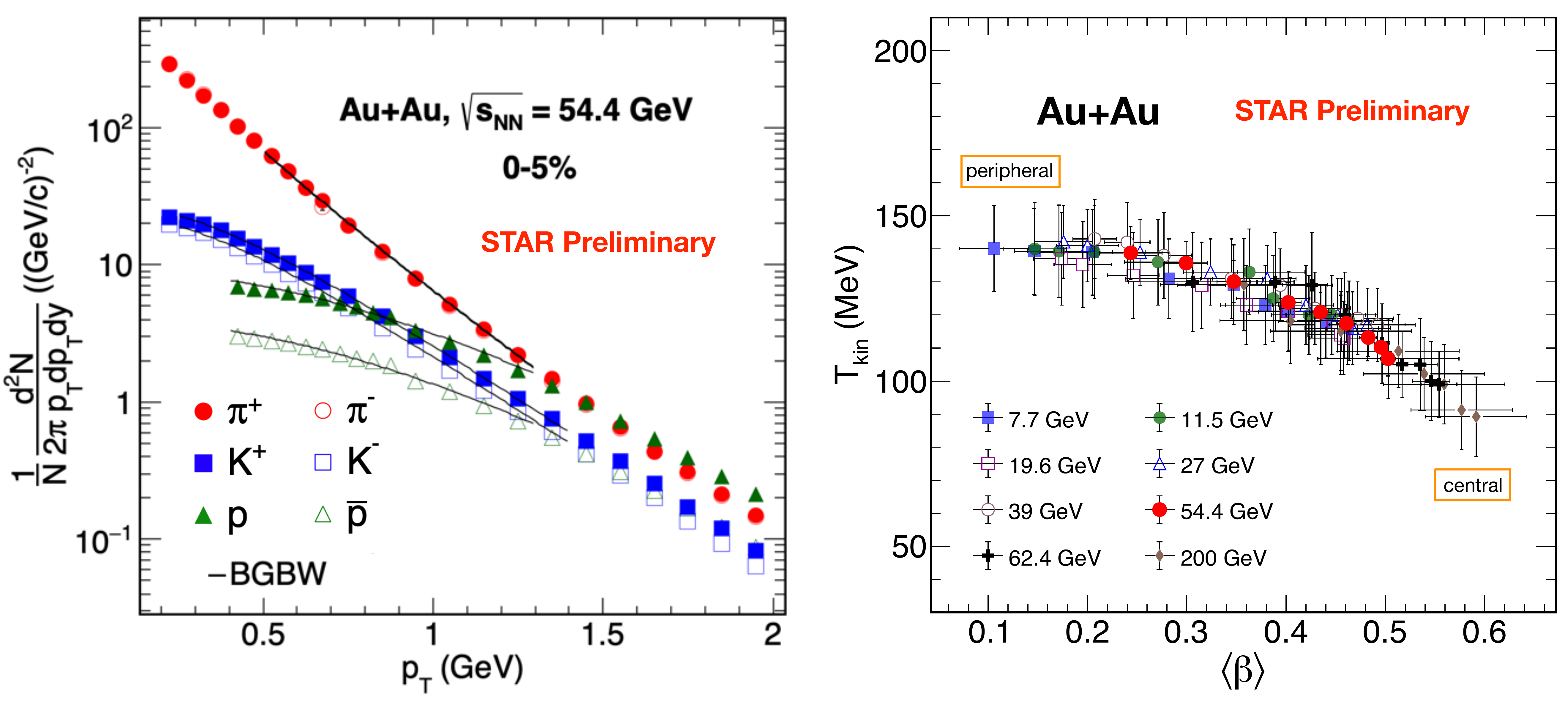}
    \caption{Left panel: Simultaneous Blast-Wave model fits to the $p_{T}$-spectra of $\pi^{±}$, $K^{±}$, $p$($\bar{p}$) in 0–5\% central Au+Au collisions at $\sqrt{s_{NN}}$ = 54.4 GeV. Right panel: Variation of $T_{kin}$ with $\langle \beta \rangle$ for different energies and collision centralities.}
    \label{fig:Fig.6}
\end{figure}
\vspace{-0.9cm}
\section{Conclusions} 
We presented transverse momentum spectra for $\pi^{\pm}$, $K^{\pm}$,  and $p$($\bar{p}$) in Au+Au collisions at $\sqrt{s_{NN}}$ = 54.4 GeV. Particle ratios from Au+Au collisions at $\sqrt{s_{NN}}$  = 54.4 GeV follow the trends from measurements at lower and higher energies. The yields of $\pi^{\pm}$, $K^{\pm}$ and $\bar{p}$ decrease with decreasing collision energy, whereas the proton yield is the highest at 7.7 GeV. Also the $p/\pi^{+}$ ratio is the largest at $\sqrt{s_{NN}}$ = 7.7 GeV and decreases with increasing energy due to the reduced baryon stopping.
%due to low baryon stopping as we go to a higher center of mass energy. 
The kinetic freeze-out temperature and $\langle \beta \rangle$ show an anti-correlation. $T_{kin}$ decreases from peripheral to central collisions indicating a short-lived fireball in peripheral collisions and $\langle \beta \rangle$ increases from peripheral to central collisions indicating a larger radial flow for central collisions.

%\printbibliography


\vspace{-0.3cm}
\begin{thebibliography}{}
\bibitem{RefC}
K. Rajagopal and F. Wilczek, arXiv hep-ph/0011333 (2000).
\bibitem{star}
K. Ackermann et al. (STAR Collaboration), Nucl. Instrum. Meth. A 499, 624 (2003).
\bibitem{bichsel}
H. Bichsel, Nucl. Instrum. Meth. A 562, 154 (2006).
%
\bibitem{RefJ}
B. I. Abelev et al. (STAR Collaboration), Phys. Rev. C 79, 34909
(2009).
\bibitem{RefD}
B. I. Abelev et al. (STAR Collaboration), Phys. Rev. C 96, 044904 (2017).
\bibitem{blastwave}
E. Schnedermann, J. Sollfrank, and U. Heinz, Phys. Rev. C 48, 2462 (1993).
\bibitem{blastwave2}
D. Teaney, J. Lauret, and E.V. Shuryak, arXiv:nucl- th/0110037.

% etc
\end{thebibliography}
\end{document}